\documentclass[aps,prd,onecolumn,groupedaddress,showpacs,nofootinbib,amssymb, preprint]{revtex4-2}
\usepackage[dvips]{graphicx}
\usepackage{amssymb}
\usepackage{amsmath}
\usepackage{graphicx,,color}
\usepackage{amsfonts}
\usepackage{bm}
\usepackage{cancel}
\usepackage{comment}
\usepackage{floatflt}
\usepackage{slashed}

\newcommand\be{\begin{equation}}
\newcommand\ee{\end{equation}}

\newcommand\e{\mathrm{e}}

\allowdisplaybreaks[4]
\begin{document}

\preprint{KEK-TH-2663, KEK-Cosmo-0362}
\title{
Non-metricity approach to Jackiw-Teitelboim gravity
}

\author{Shin'ichi~Nojiri$^{1,2}$}\email{nojiri@gravity.phys.nagoya-u.ac.jp}
\author{S.~D.~Odintsov$^{3,4}$}\email{odintsov@ice.csic.es}

\affiliation{ $^{1)}$ Theory Center, High Energy Accelerator Research Organization (KEK), Oho 1-1, Tsukuba, Ibaraki 305-0801, Japan \\
$^{2)}$ Kobayashi-Maskawa Institute for the Origin of Particles and the Universe, Nagoya University, Nagoya 464-8602, Japan \\
$^{3)}$ Institute of Space Sciences (ICE, CSIC) C. Can Magrans s/n, 08193 Barcelona, Spain \\
$^{4)}$ ICREA, Passeig Lluis Companys, 23, 08010 Barcelona, Spain
}

\begin{abstract}

We construct 2d Jackiw-Teitelboim (JT) gravity  in the framework of symmetric teleparallel gravity 
based on a non-metricity tensor. 
In symmetric teleparallel gravity, we often use the scalar quantity $Q$ composed of the bilinear terms of the non-metricity tensor 
but $Q$ is not a unique scalar quantity from the viewpoint of covariance. 
However, other scalar quantities composed of the non-metricity tensor generate ghosts in general. 
In two dimensions, because a propagating gravitational mode does not exist, 
if we consider general scalar quantities, there might appear any propagating modes although these modes are often ghosts. 
We consider a general combination of the terms given by the bi-linear of the non-metricity tensors. 
After finding the condition that the conformal form of the metric is compatible with the coincident gauge, where connections 
vanish, we determine the combination so that a solution with constant curvature exists. 
The obtained model corresponds to 2d JT gravity in the framework of the symmetric teleparallel gravity.
The conformal mode propagates as a scalar in this theory and we also consider the condition that the scalar mode is not a ghost.

\end{abstract}

\maketitle

\section{Introduction}

Recently Jackiw-Teitelboim (JT) gravity~\cite{Teitelboim:1983ux, Jackiw:1984je} has been very actively studied because this model could be 
equivalent to the Sachdev-Ye-Kitaev (SYK) model~\cite{Sachdev:1992fk, Kitaev}. 
The relation between JT gravity and SYK model has been confirmed in \cite{Jensen:2016pah, Maldacena:2016upp} 
which opens new connections between different areas of theoretical physics. 
JT gravity is a two-dimensional model where the scalar curvature is a constant. 
Since the scalar curvature $R$ is a total derivative or topological density in two dimensions, Einstein gravity with a cosmological constant $\Lambda$ 
becomes inconsistent. 
Therefore Einstein gravity cannot describe JT gravity. 
An action realising JT gravity is given by introducing an auxiliary scalar field $\phi$, 
\begin{align}
\label{WFR7B}
S=\frac{1}{2\kappa^2} \int dx^2 \sqrt{-g} \, \phi \left( R - \Lambda \right)\, . 
\end{align}
By the variation of the action with respect to $\phi$, the obtained equation shows that the scalar curvature is a constant $R=\Lambda$. 
In Ref.~\cite{Nojiri:2022mfi} (see also \cite{Nojiri:2023saw}), it has been shown that the action (\ref{WFR7B}) can be rewritten by the 
$F(R)$ gravity~\cite{Capozziello:2002rd, Nojiri:2003ft} 
(see \cite{Capozziello:2009nq, Faraoni:2010pgm, Capozziello:2011et, Nojiri:2010wj, Nojiri:2017ncd} for reviews of general modified gravity 
theories). 

Recently in \cite{Boehmer:2024gxo}, it has been shown that JT gravity can be formulated by teleparallel gravity using a torsion tensor. 
In teleparallel gravity, usually, one uses the torsion scalar $T$~\cite{Hehl:1976kj, Hayashi:1979qx, Bengochea:2008gz, Linder:2010py}, 
whose difference from the scalar curvature in Einstein gravity is a total derivative. 
Note, however, that the torsion scalar is not a unique scalar composed of the torsion tensor. 
In \cite{Boehmer:2024gxo}, another scalar quantity different from the torsion scalar $T$ is used to reproduce JT gravity. 

In general, we may construct a gravity theory based on the gauge theory of the Poincar\'{e} algebra. 
The Poincar\'{e} algebra is composed of the generators of the Lorentz transformation and the translation. 
By requiring the curvature of the translation to vanish, the curvature of the Lorentz transformation gives the curvature of Einstein's gravity. 
The curvature of the translation corresponds to the torsion tensor. 
In \cite{Boehmer:2024gxo}, the action is chosen to be a square of the curvature corresponding to the translation, that is, field strength, 
which is an analogue of the electrodynamics or non-abelian gauge theory. 

In this paper, we construct JT gravity in the framework of symmetric teleparallel gravity 
using a non-metricity tensor~\cite{Nester:1998mp, BeltranJimenez:2018vdo, Runkla:2018xrv, Capozziello:2022tvv}, 
where both the curvature and the torsion vanish. 
In symmetric teleparallel gravity, we often use the scalar quantity $Q$ composed of quantities called non-metricity tensors. 
 From the viewpoint of covariance, $Q$ is not a unique scalar quantity although the difference between $Q$ and the scalar curvature in Einstein's gravity is a total derivative 
as in the case of the torsion scalar. 
If we consider the perturbation in the flat Minkowski background, $Q$ reduces to the Lagrangian density of the gravitational wave. 
This shows that $Q$ could be a unique scalar quantity composed of bilinear of the non-metricity tensors 
so that the action linear to $Q$ gives only the massless spin-two mode but does not include extra modes, which are often ghosts. 
In two dimensions, the situation is rather special because there is no propagating mode in Einstein's gravity. 
Therefore if one considers more general scalar quantities composed of the non-metricity tensors, there could be any propagating modes. 
Such modes are often ghosts, whose kinetic energy is not bounded below. 
If we change the signature of the action, the mode becomes canonical and the kinetic energy is bounded below. 
In the higher dimensions, if we change the signature of the action, the standard gravitational wave becomes a ghost and the kinetic energy becomes unbounded below. 
In the two dimensions, because the propagating massless spin-two mode does not exist, even if we change the signature of the action, there might not appear any ghost. 
{ 
We should note that the symmetric teleparallel gravity in two dimensions was first considered in \cite{Adak:2006xx} 
and the general scalar combinations composed of the non-metricity tensors were also investigated in \cite{Adak:2005cd, Doyran:2023zhk}. 

}

A problem proper to the symmetric teleparallel gravity is the treatment of the connections. 
In symmetric teleparallel gravity, the connections are degrees of freedom independent from metric although the connections 
are constrained by conditions that require the curvature and the torsion to vanish. 
The structure of the constraints is very complicated and the number of degrees of freedom is not clear even off-shell. 
Therefore it is very difficult to count the number of physical degrees of freedom. 
We often use the coincident gauge condition 
{ 
invented in \cite{Adak:2006rx, Adak:2008gd}, 
}
where the connections vanish. 
This gauge condition makes the equations simple but the structure of the constraints is still not clear and 
because the coincident gauge fixes the general covariance, the solutions severely depend on the coordinate choice and it becomes difficult to find the coordinate system 
consistent with the coincident gauge. 
In order to clarify the off-shell degrees of freedom, in \cite{Nojiri:2024zab}, the metric and four scalar fields are chosen to be independent fields off-shell. 
Although the number of physical degrees of freedom is still not clear, all the equations are given in a closed form. 

In this paper, we start with a general combination composed of the bi-linear of the non-metricity tensors. 
We assume the conformal form of the metric in two spacetime dimensions, 
\begin{align}
\label{conf}
g_{\mu\nu} = \e^\sigma \eta_{\mu\nu}\, , \quad \mbox{that is,}\quad 
g_{00}=-\e^\sigma\, , \quad g_{11}=\e^\sigma\, , \quad g_{01}=g_{10}=0\, .
\end{align}
As this fixes the gauge degrees of freedom in the general covariance, we cannot assume the coincident gauge condition in addition to 
the condition (\ref{conf}). 
Hence the coincident gauge condition makes the equations simpler, however, we consider the condition that the coincident gauge condition is compatible with 
the assumption (\ref{conf}) on the metric. 
We find that there exist models which satisfy the condition. 
After that, we consider the conditions that the model has a solution of the configuration where the scalar curvature is a constant as in JT gravity. 
Finally, we find a model corresponding to JT gravity in the framework of symmetric teleparallel gravity. 

\section{Brief review of $f(Q)$ gravity\label{SecII}}

The general affine connection on a manifold that is both parallelizable and differentiable can be expressed as follows:
\begin{align}
\label{affine}
{\Gamma^\sigma}_{\mu \nu}= {{\tilde \Gamma}^\sigma}_{\mu \nu} + K^\sigma_{\;\mu \nu} + L^\sigma_{\;\mu \nu}\,.
\end{align}
Here $\tilde \Gamma^\sigma_{\;\mu \nu}$ is the Levi-Civita connection given by the metric,
\begin{align}
\label{Levi-Civita}
{{\tilde\Gamma}^\sigma}_{\mu \nu} = \frac{1}{2} g^{\sigma \rho} \left( \partial_\mu g_{\rho \nu} + \partial_\nu g_{\rho \mu}- \partial_\rho g_{\mu \nu}\right)\,.
\end{align}
Furthermore, ${K^\sigma}_{\mu \nu}$ represents the contortion which is defined by using the torsion tensor
${T^\sigma}_{\mu \nu}={\Gamma^\sigma}_{\mu \nu} - {\Gamma^\sigma}_{\nu \mu}$ as follows, 
\begin{align}
\label{contortion}
{K^\sigma}_{\mu \nu}= \frac{1}{2} \left( {T^\sigma}_{\mu \nu} + T^{\ \sigma}_{\mu\ \nu} + T^{\ \sigma}_{\nu\ \mu} \right) \, .
\end{align}
Finally, ${L^\sigma}_{\mu \nu}$ denotes the deformation and is expressed as follows:
\begin{align}
\label{deformation}
{L^\sigma}_{\mu \nu}= \frac{1}{2} \left( Q^\sigma_{\;\mu \nu} - Q^{\ \sigma}_{\mu\ \nu} - Q^{\ \sigma}_{\nu\ \mu} \right)\,.
\end{align}
Here ${Q^\sigma}_{\mu \nu}$ represents the non-metricity tensor expressed as,
\begin{align}
\label{non-metricity}
Q_{\sigma \mu \nu}= \nabla_\sigma g_{\mu \nu}= \partial_\sigma g_{\mu \nu} - {\Gamma^\rho}_{\sigma \mu } g_{\nu \rho} - {\Gamma^\rho}_{\sigma \nu } g_{\mu \rho } \,.
\end{align}
We use $Q_{\sigma \mu \nu}$ in order to construct the $f(Q)$ gravity. 

In the following, we consider the case without torsion, therefore we assume ${\Gamma^\sigma}_{\mu\nu} = {\Gamma^\sigma}_{\nu\mu}$. 
Symmetric teleparallel theories of gravity are obtained by requiring the Riemann tensor to vanish, 
\begin{align}
\label{curvatures}
R^\lambda_{\ \mu\rho\nu} \equiv \Gamma^\lambda_{\mu\nu,\rho} -\Gamma^\lambda_{\mu\rho,\nu} + \Gamma^\eta_{\mu\nu}\Gamma^\lambda_{\rho\eta}
 - \Gamma^\eta_{\mu\rho}\Gamma^\lambda_{\nu\eta} =0 \, ,
\end{align}
whose solution is given by using four fields $\xi^a$ $\left( a = 0,1,2,3 \right)$ as follows, 
\begin{align}
\label{G1B}
{\Gamma^\rho}_{\mu\nu}=\frac{\partial x^\rho}{\partial \xi^a} \partial_\mu \partial_\nu \xi^a \, .
\end{align}
As we show soon, $\xi^a$'s should be scalar fields and we may regard $e^a_\mu\equiv \partial_\mu \xi^a$'s as vierbein fields. 
There is a gauge symmetry of the general coordinate transformation, so we often choose the gauge condition ${\Gamma^\rho}_{\mu\nu}=0$, which is called 
the coincident gauge and can be realised by choosing 
\begin{align}
\label{Cgauge}
\xi^a=x^a \, . 
\end{align}
The gauge condition, however, often contradicts the existence of the FLRW universe and the spherically symmetric spacetime. 

When one considers the infinitesimal transformation, $\xi^a \to \xi^a + \delta \xi^a$, one finds
\begin{align}
\label{G1}
\Gamma^\rho_{\mu\nu} \to \Gamma^\rho_{\mu\nu} + \delta \Gamma^\rho_{\mu\nu}
\equiv \Gamma^\rho_{\mu\nu} - \frac{\partial x^\rho}{\partial \xi^a} \partial_\sigma \delta\xi^a\frac{\partial x^\sigma}{\partial \xi^b}\partial_\mu \partial_\nu \xi^b
+ \frac{\partial x^\rho}{\partial \xi^a} \partial_\mu \partial_\nu \delta \xi^a \, , 
\end{align}
which can be used later to find the field equations. 
By regarding $\xi^a$'s as scalar fields, under the coordinate transformation $x^\mu\to x^\mu + \epsilon^\mu$ with infinitesimally small functions $\epsilon^\mu$, 
we find $\delta\xi^a = \epsilon^\mu \partial_\mu \xi^a$ and therefore 
\begin{align}
\label{G1GCT}
\delta \Gamma^\rho_{\mu\nu}
= \epsilon^\sigma \partial_\sigma \Gamma^\rho_{\mu\nu} - \partial_\sigma \epsilon^\rho \Gamma^\sigma_{\mu\nu} 
+ \partial_\mu \epsilon^\eta \Gamma^\rho_{\eta\nu} + \partial_\nu \epsilon^\eta \Gamma^\rho_{\mu\eta} 
+ \partial_\mu \partial_\nu \epsilon^\rho \, , 
\end{align}
Here the last term is nothing but the inhomogeneous term, which guarantees the general covariance of the covariant derivative. 
This also shows that $\xi^a$'s are not vector fields but scalar fields. 

In the symmetric teleparallel theory, we use the non-metricity tensor in (\ref{non-metricity}) and the scalar $Q$ of the non-metricity defined as follows, 
\begin{align}
\label{non-m scalar}
Q\equiv g^{\mu \nu} \left( {L^\alpha}_{\beta \nu}{L^\beta}_{\mu \alpha} - {L^\beta}_{\alpha \beta} {L^\alpha}_{\mu \nu} \right)
 -Q_{\sigma \mu \nu} P^{\sigma \mu \nu} \,.
\end{align}
Here $P^{\sigma \mu \nu}$ is defined by,
\begin{align}
\label{non-m conjugate}
{P^\sigma}_{\mu \nu} \equiv &\, \frac{1}{4} \left\{ - {Q^\sigma}_{\mu \nu} + Q^{\ \sigma}_{\mu\ \nu} + Q^{\ \sigma}_{\nu\ \mu}
+ Q^\sigma g_{\mu \nu}- \tilde{Q}^\sigma g_{\mu \nu} - \frac{1}{2} \left( {\delta^\sigma}_\mu Q_\nu + {\delta^\sigma}_\nu Q_\mu \right) \right\}\, ,
\end{align}
and $Q_\sigma$ and $\tilde{Q}_\sigma$ are defined as
$Q_\sigma \equiv Q^{\ \mu}_{\sigma\ \mu}$ and $\tilde{Q}_\sigma=Q^\mu_{\ \sigma \mu}$. 
Then one obtains, 
\begin{align}
\label{Q}
Q=&\, - \frac{1}{4} g^{\alpha\mu} g^{\beta\nu} g^{\gamma\rho} \nabla_\alpha g_{\beta\gamma} \nabla_\mu g_{\nu\rho}
+ \frac{1}{2} g^{\alpha\mu} g^{\beta\nu} g^{\gamma\rho} \nabla_\alpha g_{\beta\gamma} \nabla_\rho g_{\nu\mu}
+ \frac{1}{4} g^{\alpha\mu} g^{\beta\gamma} g^{\nu\rho} \nabla_\alpha g_{\beta\gamma} \nabla_\mu g_{\nu\rho} \nonumber \\
&\, - \frac{1}{2} g^{\alpha\mu} g^{\beta\gamma} g^{\nu\rho} \nabla_\alpha g_{\beta\gamma} \nabla_\nu g_{\mu\rho} \, .
\end{align}
The difference between $Q$ and the scalar curvature $\tilde R$ of Einstein gravity given by the Levi-Civita connection is the total derivative, 
$\tilde R = Q - {\tilde\nabla}_\alpha\left(Q^{\alpha}-\tilde{Q}^{\alpha}\right)$. 
Here ${\tilde\nabla}_\alpha$ is a covariant derivative defined by the Levi-Civita connection. 
Therefore the action linear to $Q$ is equivalent to the Einstein-Hilbert action in Einstein gravity. 

On the other hand, the action of $f(Q)$ gravity with a function $f(Q)$ of $Q$, that is, 
$S=\int d^4 x \sqrt{-g} f(Q)$, 
is different from the action of $f(R)$ gravity. 
In the following, we regard the metric $g_{\mu\nu}$ and $\xi^a$ as independent fields as proposed in \cite{Nojiri:2024zab}. 

\section{Curvature degrees of freedom }

The Lagrange multiplier fields are used to put constraints that make the curvature and the torsion vanish. 
The counting of the degrees of freedom is very tedious and the situation makes the Hamiltonian analysis very complicated. 
That is a motivation why we start the model by the metric $g_{\mu\nu}$ and scalar fields $\xi^a$ as independent fields~\cite{Nojiri:2024zab}.
In this section, we explain the situation. 

By choosing that the indices $\nu$ and $\rho$ are symmetric under the exchange in the connection, $\Gamma^\mu_{\nu\rho}=\Gamma^\mu_{\rho\nu}$, 
the torsion can be excluded rather easily. 
In general, even if there is no torsion, the connection $\Gamma^\mu_{\nu\rho}$ has $\frac{D^2 \left(D+1\right)}{2}$ components. 
Here $D$ is the dimension of the spacetime. 
This is because $\mu$ is independent and the indices $\nu$ and $\rho$ are symmetric under the exchange. 

The Riemann curvature $R^\lambda_{\ \mu\rho\nu}$ defined by (\ref{curvatures}) has the following properties, 
$R^\lambda_{\ \mu\rho\nu} = - R^\lambda_{\ \mu\nu\rho}$, 
$0 = R^\lambda_{\ \mu\rho\nu} + R^\lambda_{\ \rho\nu\mu} + R^\lambda_{\ \nu\mu\rho}$. 
Not as in the case of Einstein gravity, there is no symmetry in $\lambda$ and $\mu$. 
Then in two dimensions the possible and independent combinations of $( \mu, \rho, \nu )$ is given by 
$( \mu, \rho, \nu ) = (0, 0, 1)$, $(1, 0, 1)$. 
On the other hand, in three dimensions, we find $8$ independent components, 
$( \mu, \rho, \nu ) =(0,0,1)$, $(0,0,2)$, $(0,1,2)$, $(1,0,1)$, $(1,0,2)$, $(1,1,2)$, $(2,0,2)$, $(2,1,2)$, 
and in four dimensions, we find $20$ independent components, 
$( \mu, \rho, \nu ) = (0,0,1)$, $(0,0,2)$, $(0,0,3)$, $(0,1,2)$, $(0,1,3)$, $(0,2,3)$, $(1,0,1)$, $(1,0,2)$, $(1,0,3)$, $(1,1,2)$, $(1,1,3)$, $(1,2,3)$, $(2,0,2)$, $(2,0,3)$, 
$(2,1,2)$, $(2,1,3)$, $(2,2,3)$, $(3,0,3)$, $(3,1,3)$, $(3,2,3)$. 
In general $d$ dimensions $( \mu, \rho, \nu )$ has $\frac{2D(D-1)}{2!} + 2\frac{D(D-1)(D-2)}{3!} = \frac{(D+1) D (D-1)}{3}$ independent components and 
the Riemann tensor has $\frac{(D+1) D^2 (D-1)}{3}$ independent components. 

Then in two dimensions, the connection $\Gamma^\mu_{\nu\rho}$ has $6$ components 
and the Riemann tensor $R^\lambda_{\ \mu\rho\nu}$ has $2\times 2=4$ components. 
Therefore when we impose the constraint, 
\begin{align}
\label{cons1}
0= R^\lambda_{\ \mu\rho\nu} \, ,
\end{align}
there remains two components and therefore the solution of (\ref{cons1}) is given by two scalar fields $\xi^a$ $\left(a=0,1\right)$, 
as in (\ref{G1B}). 
In three dimensions, however, the connection $\Gamma^\mu_{\nu\rho}$ has $18$ components but 
the Riemann tensor $R^\lambda_{\ \mu\rho\nu}$ has $24$ components. 
Therefore there are $6$ remaining degrees of freedom. 
Furthermore in four dimensions, the connection 
 $\Gamma^\mu_{\nu\rho}$ has $40$ components but 
the Riemann tensor $R^\lambda_{\ \mu\rho\nu}$ has $80$ components. 
Then naively the number of the constraints is larger than that of the degrees of freedom of the connection. 
Although Eq.~(\ref{G1B}) is a solution in general $d$ dimensions, we do not know the real number of degrees of freedom of the connection except $d=2$. 

\section{New symmetric teleparallel gravity}

The scalar $Q$ (\ref{Q}) is chosen so that the difference from the scalar curvature $\mathcal{R}$ of Einstein gravity becomes a total derivative. 
 From the viewpoint of the general covariance, this is not a unique choice because $\nabla_\rho g_{\mu\nu}$ is a tensor. 
For example, we may consider a more general combination as follows, 
\begin{align}
\label{AQ}
A_Q=&\, a g^{\alpha\mu} g^{\beta\nu} g^{\gamma\rho} \nabla_\alpha g_{\beta\gamma} \nabla_\mu g_{\nu\rho}
+ b g^{\alpha\mu} g^{\beta\nu} g^{\gamma\rho} \nabla_\alpha g_{\beta\gamma} \nabla_\rho g_{\nu\mu}
+ c g^{\alpha\mu} g^{\beta\gamma} g^{\nu\rho} \nabla_\alpha g_{\beta\gamma} \nabla_\mu g_{\nu\rho} \nonumber \\
&\, +d g^{\alpha\mu} g^{\beta\gamma} g^{\nu\rho} \nabla_\alpha g_{\beta\gamma} \nabla_\nu g_{\mu\rho} \, ,
\end{align}
with constant parameters, $a$, $b$, $c$, and $d$, 
and consider the following action, 
\begin{align}
\label{SAQ}
S_{A_Q} \equiv \int \sqrt{-g} A_Q\, ,
\end{align}
We consider the above action (\ref{SAQ}), hereafter. 

Note that the combination of $Q$ in (\ref{Q}) is special. 
Starting from the following action, 
\begin{align}
\label{QAction}
S_Q = \frac{1}{8\kappa^2} \int d^4 x \sqrt{-g} Q \, .
\end{align}
when we consider the perturbation $g_{\mu\nu} = \eta_{\mu\nu} + h_{\mu\nu}$ of the flat Minkowski background whose metric is given by $\eta_{\mu\nu}$, 
the action~(\ref{QAction}) reduces to the action of the gravitational wave
\begin{align}
\label{fltGW}
S_Q \sim \frac{1}{8\kappa^2} \int d^4 x &\, \left\{ 
- \frac{1}{4} \eta^{\alpha\mu} \eta^{\beta\nu} \eta^{\gamma\rho} \partial_\alpha h_{\beta\gamma} \partial_\mu h_{\nu\rho}
+ \frac{1}{2} \eta^{\alpha\mu} \eta^{\beta\nu} \eta^{\gamma\rho} \partial_\alpha h_{\beta\gamma} \partial_\rho h_{\nu\mu} \right. \nonumber \\
&\, \left. + \frac{1}{4} \eta^{\alpha\mu} \eta^{\beta\gamma} \eta^{\nu\rho} \partial_\alpha h_{\beta\gamma} \partial_\mu h_{\nu\rho}
 - \frac{1}{2} \eta^{\alpha\mu} \eta^{\beta\gamma} \eta^{\nu\rho} \partial_\alpha h_{\beta\gamma} \partial_\nu h_{\mu\rho} \right\} \, 
\end{align}
Similarly in the case of the action~(\ref{SAQ}), one obtains 
\begin{align}
\label{AQGW}
S_{A_Q} \sim \frac{1}{8\kappa^2} \int d^4 x &\, \left\{ 
a \eta^{\alpha\mu} \eta^{\beta\nu} \eta^{\gamma\rho} \partial_\alpha h_{\beta\gamma} \partial_\mu h_{\nu\rho}
+ b \eta^{\alpha\mu} \eta^{\beta\nu} \eta^{\gamma\rho} \partial_\alpha h_{\beta\gamma} \partial_\rho h_{\nu\mu}
\right. \nonumber \\
&\, \left. + c \eta^{\alpha\mu} \eta^{\beta\gamma} \eta^{\nu\rho} \partial_\alpha h_{\beta\gamma} \partial_\mu h_{\nu\rho} 
+d \eta^{\alpha\mu} \eta^{\beta\gamma} \eta^{\nu\rho} \partial_\alpha h_{\beta\gamma} \partial_\nu h_{\mu\rho} \right\} \, 
\end{align}
In order to kill the unphysical modes, we require that 
the coefficients of $\left( \partial_0 h_{00} \right)^2$ and $\left( \partial_0 h_{i0}\right)^2$ should vanish 
and we obtain conditions $0= - \left(a+b+c+d\right)$ and $0=2a + b$, 
which are satisfied in $Q$ or the action (\ref{QAction}). 
We rewrite the conditions as $b= - 2a$ and $d= a - c$. 
When we parametrise $a=- \frac{\alpha}{4}$ and $c=\frac{\alpha\gamma}{4}$, 
by further requiring $\gamma=1$, other unphysical modes are excluded. 
In two dimensions, no mode corresponds to the gravitational wave. 
Furthermore, in Einstein gravity, any mode does not propagate. 
In order to obtain a model including non-trivial propagating mode in two dimensions, 
we consider the general action~(\ref{SAQ}). 

The variation of the action~(\ref{SAQ}) with respect to the metric $g_{\mu\nu}$ gives 
\begin{align}
\label{eq1}
\mathcal{G}_{\mu\nu} \equiv &\, \frac{1}{\sqrt{-g}} g_{\mu\rho} g_{\nu\sigma}\frac{\delta S_{A_Q}}{\delta g_{\rho\sigma}} \nonumber \\
=&\, \frac{1}{2} g_{\mu\nu} \left\{ a g^{\alpha\xi} g^{\beta\zeta} g^{\gamma\rho} \nabla_\alpha g_{\beta\gamma} \nabla_\xi g_{\zeta\rho}
+ b g^{\alpha\xi} g^{\beta\zeta} g^{\gamma\rho} \nabla_\alpha g_{\beta\gamma} \nabla_\rho g_{\zeta\xi}
+ c g^{\alpha\xi} g^{\beta\gamma} g^{\zeta\rho} \nabla_\alpha g_{\beta\gamma} \nabla_\xi g_{\zeta\rho} 
\right. \nonumber \\
&\, \left. 
+d g^{\alpha\xi} g^{\beta\gamma} g^{\zeta\rho} \nabla_\alpha g_{\beta\gamma} \nabla_\zeta g_{\xi\rho} \right\} \nonumber \\
&\, - g^{\alpha\beta} g^{\gamma\rho} \left\{ a \nabla_\mu g_{\alpha\gamma} \nabla_\nu g_{\beta\rho}
+ 2 a \nabla_\alpha g_{\mu\gamma} \nabla_\beta g_{\nu\rho} \right. \nonumber \\
&\, + b \left( \nabla_\mu g_{\alpha\gamma} \nabla_\rho g_{\beta\nu}
+ \nabla_\nu g_{\alpha\gamma} \nabla_\rho g_{\beta\mu} \right)
+ b \nabla_\alpha g_{\mu\gamma} \nabla_\rho g_{\nu\beta}
+ c \nabla_\mu g_{\alpha\beta} \nabla_\nu g_{\gamma\rho}
+ 2c \nabla_\alpha g_{\mu\nu} \nabla_\beta g_{\gamma\rho} \nonumber \\
&\, \left. + \frac{d}{2} \left( \nabla_\mu g_{\alpha\beta} \nabla_\gamma g_{\nu\rho} + \nabla_\nu g_{\alpha\beta} \nabla_\gamma g_{\mu\rho} \right)
+ d \nabla_\alpha g_{\mu\nu} \nabla_\gamma g_{\beta\rho}
+ \frac{d}{2} \left(\nabla_\alpha g_{\gamma\rho} \nabla_\mu g_{\beta\nu} + \nabla_\alpha g_{\gamma\rho} \nabla_\nu g_{\beta\mu} \right)
\right\} \nonumber \\
&\, - \frac{g_{\mu\rho} g_{\nu\sigma}}{\sqrt{-g}} \partial_\alpha \left[ \sqrt{-g} \left\{ 2a g^{\alpha\beta} g^{\gamma\rho} g^{\sigma\tau} \nabla_\beta g_{\gamma\tau}
+ b g^{\alpha\beta} g^{\gamma\rho} g^{\sigma\tau} \left( \nabla_\tau g_{\gamma\beta} + \nabla_\gamma g_{\tau\beta} \right) \right. \right. \nonumber \\
&\, \left. \left. + 2c g^{\alpha\beta} g^{\rho\sigma} g^{\gamma\tau} \nabla_\beta g_{\gamma\tau}
+ d g^{\alpha\beta} g^{\rho\sigma} g^{\gamma\tau} \nabla_\gamma g_{\beta\tau}
+ \frac{d}{2} \left( g^{\alpha\sigma} g^{\gamma\tau} g^{\beta\rho} + g^{\alpha\rho} g^{\gamma\tau} g^{\beta\sigma}
\right) \nabla_\beta g_{\gamma\tau}
\right\} \right] \nonumber \\
&\, - \frac{g_{\mu\rho} g_{\nu\sigma}}{\sqrt{-g}} {\Gamma^\rho}_{\alpha\eta} \left[ \sqrt{-g} \left\{ 2a g^{\alpha\beta} g^{\gamma\eta} g^{\sigma\tau} \nabla_\beta g_{\gamma\tau}
+ b g^{\alpha\beta} g^{\gamma\eta} g^{\sigma\tau} \left( \nabla_\tau g_{\gamma\beta} + \nabla_\gamma g_{\tau\beta} \right) \right. \right. \nonumber \\
&\, \left. \left. + c g^{\alpha\beta} g^{\eta\sigma} g^{\gamma\tau} \nabla_\beta g_{\gamma\tau}
+ d g^{\alpha\beta} g^{\eta\sigma} g^{\gamma\tau} \nabla_\gamma g_{\beta\tau}
+ \frac{d}{2} \left( g^{\alpha\sigma} g^{\gamma\tau} g^{\beta\eta} + g^{\alpha\eta} g^{\gamma\tau} g^{\beta\sigma}
\right) \nabla_\beta g_{\gamma\tau}
\right\} \right] \nonumber \\
&\, - \frac{g_{\mu\rho} g_{\nu\sigma}}{\sqrt{-g}} {\Gamma^\sigma}_{\alpha\eta} \left[ \sqrt{-g} \left\{ 2a g^{\alpha\beta} g^{\gamma\rho} g^{\eta\tau} \nabla_\beta g_{\gamma\tau}
+ b g^{\alpha\beta} g^{\gamma\rho} g^{\eta\tau} \left( \nabla_\tau g_{\gamma\beta} + \nabla_\gamma g_{\tau\beta} \right) \right. \right. \nonumber \\
&\, \left. \left. + c g^{\alpha\beta} g^{\rho\eta} g^{\gamma\tau} \nabla_\beta g_{\gamma\tau}
+ d g^{\alpha\beta} g^{\rho\eta} g^{\gamma\tau} \nabla_\gamma g_{\beta\tau}
+ \frac{d}{2} \left( g^{\alpha\eta} g^{\gamma\tau} g^{\beta\rho} + g^{\alpha\rho} g^{\gamma\tau} g^{\beta\eta}
\right) \nabla_\beta g_{\gamma\tau}
\right\} \right] \, ,
\end{align}
 Under the variation with respect to the connection ${\Gamma^\eta}_{\xi\eta}$, we find 
\begin{align}
\label{eq2}
\mathcal{H}^{\xi\zeta}_\eta \equiv&\, \frac{1}{\sqrt{-g}} \frac{\delta S_{A_Q}}{\delta {\Gamma^\eta}_{\xi\zeta}} 
= - \Bigl( 
2a g^{\xi\rho} g^{\zeta\nu} g^{\gamma\mu} 
+ a g^{\xi\rho} g^{\gamma\nu} g^{\zeta\mu} + a g^{\zeta\rho} g^{\gamma\nu} g^{\xi\mu} 
+ 2b g^{\xi\mu} g^{\zeta\nu} g^{\gamma\rho} \nonumber \\
&\, + 2b g^{\zeta\mu} g^{\xi\nu} g^{\gamma\rho} 
+ 2c g^{\xi\mu} g^{\zeta\gamma} g^{\nu\rho} 
+ 2c g^{\zeta\mu} g^{\xi\gamma} g^{\nu\rho} 
+ d g^{\xi\nu} g^{\zeta\gamma} g^{\mu\rho} + d g^{\zeta\nu} g^{\xi\gamma} g^{\mu\rho} \nonumber \\
&\, + \frac{d}{2} g^{\mu\zeta} g^{\nu\rho} g^{\xi\gamma} 
+ \frac{d}{2} g^{\mu\gamma} g^{\nu\rho} g^{\xi\zeta} 
\Bigr) g_{\eta\gamma} \nabla_\mu g_{\nu\rho} \, .
\end{align}
Therefore the variation with respect to $\xi^a$ in (\ref{G1B}) gives the following equation, 
\begin{align}
\label{eq3}
X_a \equiv \frac{1}{\sqrt{-g}} \frac{\delta S}{\delta \xi^a} 
= \partial_\sigma
\left\{ \frac{\partial x^\eta}{\partial \xi^a} \frac{\partial x^\sigma}{\partial \xi^b}\partial_\xi \partial_\zeta \xi^b 
\left( \sqrt{-g} \mathcal{H}^{\xi\zeta}_\eta \right) \right\}
+ \partial_\xi \partial_\zeta \left\{ \frac{\partial x^\eta}{\partial \xi^a} \left( \sqrt{-g} \mathcal{H}^{\xi\zeta}_\eta \right) \right\} \, .
\end{align}
In the following, we consider the model in two dimensions by using the above equations.

\section{Two dimensional case}

In two dimensions, the metric has three components and there are two gauge symmetries corresponding to the diffeomorphism. 
Therefore we often delete the two components of the metric and use the conformal gauge, where the metric is given by (\ref{conf}). 
Then one finds $\nabla_\mu g_{\nu\rho} = \e^\sigma \left( \partial_\mu \sigma \eta_{\nu\rho} - {\Gamma^\tau}_{\mu\nu} \eta_{\tau\rho} 
 - {\Gamma^\tau}_{\mu\rho} \eta_{\nu\tau} \right)$, 
and 
\begin{align}
\label{eq2B}
\mathcal{H}^{\xi\zeta}_\eta 
=&\, - \e^{-\sigma}\left[ \left(2a + d \right) \eta^{\xi\zeta} \partial_\eta \sigma 
+ \left( a + 2b + 4c + \frac{3d}{2} \right) \left( \delta^\zeta_{\ \eta} \eta^{\xi\mu} \partial_\mu \sigma + \delta^\xi_{\ \eta} \eta^{\zeta\mu} \partial_\mu \sigma \right)
\right. \nonumber \\
&\, + \left( - 3 a - 2b \right) \left( \eta^{\zeta\nu} {\Gamma^\xi}_{\eta\nu} + \eta^{\xi\nu} {\Gamma^\zeta}_{\eta\nu} \right) 
+ \left( - 2a - 4b \right) \eta^{\xi\rho} \eta^{\zeta\mu} {\Gamma^\eta}_{\mu\rho} \nonumber \\
&\, \left. + \left( - 4c - \frac{3d}{2}\right) \left( \eta^{\xi\mu} \delta^\zeta_{\ \eta} {\Gamma^\tau}_{\mu\tau} + \eta^{\zeta\mu} \delta^\xi_{\ \eta} {\Gamma^\tau}_{\mu\tau} \right) 
 - d \left( \delta^\zeta_{\ \eta} \eta^{\mu\nu} {\Gamma^\xi}_{\mu\nu} + \delta^\xi_{\ \eta} \eta^{\mu\nu} {\Gamma^\zeta}_{\mu\nu} \right) 
 - d \eta^{\xi\zeta} {\Gamma^\tau}_{\eta\tau} \right] \, .
\end{align}
Let us now consider the case of the coincident gauge (\ref{Cgauge}), which results in ${\Gamma^\mu}_{\nu\rho}=0$, 
is compatible with the conformal gauge (\ref{conf}). 
Then Eqs.~(\ref{eq2}) and (\ref{eq3}) give $X_\eta =- 4 \left( a + b + 2c + d \right) \eta^{\xi\zeta} \partial_\xi \partial_\zeta \partial_\eta \sigma$. 
Therefore when we require $X_\eta=0$, we find that in the case, 
\begin{align}
\label{abcd}
0 = a + b + 2c + d \, ,
\end{align}
the coincident gauge (\ref{Cgauge}) can be consistently imposed. 
Under the coincident gauge (\ref{Cgauge}), Eq.~(\ref{eq1}) gives, 
\begin{align}
\label{eq1abcdC}
\mathcal{G}_{\mu\nu} =&\, \left[ \left( 3a + \frac{b}{2} + 6c + 2 d \right) \eta_{\mu\nu} \eta^{\xi\zeta} \partial_\xi \sigma \partial_\zeta \sigma 
+ \left( - 4a - b - 8c - 2d \right) \partial_\mu \sigma \partial_\nu \sigma \right. \nonumber \\
&\, \left. + \left( - 2a - 4c - d \right) \eta_{\mu\nu} \eta^{\alpha\beta} \partial_\alpha \partial_\beta \sigma 
+ 2 \left( - b - d \right) \partial_\mu \partial_\nu \sigma \right] \, . 
\end{align}
We may now construct an analogue of JT gravity in two spacetime dimensions in the framework of symmetric teleparallel gravity. 
The curvature is a constant in JT gravity. Let us consider the curvature in Einstein gravity under the coordinate system (\ref{conf}). 

The Levi-Civita connection for the coordinate system (\ref{conf}) is given by ${{\tilde \Gamma}^\mu}_{\nu\rho} 
= \frac{1}{2} \left( \delta^\mu_{\ \rho} \partial_\nu \sigma + \delta^\mu_{\ \nu} \partial_\rho \sigma - \eta_{\nu\rho} \eta^{\mu\xi} \partial_\xi \sigma \right)$ 
and the scalar curvature $\tilde R$ in Einstein gravity has the form $\tilde R = \eta^{\mu\nu} \partial_\mu \partial_\nu \left( \e^{-\sigma} \right)$ 
If $\sigma$ in the metric is given by 
\begin{align}
\label{esigma}
\e^{-\sigma} = \frac{1}{2l^2} \left( - t^2 + x^2 \right) = \frac{\eta_{\mu\nu} x^\mu x^\nu }{2l^2} \, , 
\end{align}
we find the scalar curvature is a constant 
\begin{align}
\label{tR}
\tilde R = \frac{2}{l^2} \, .
\end{align}

Because we have the following for (\ref{esigma}), $\sigma = - \ln \frac{\eta_{\mu\nu} x^\mu x^\nu }{2l^2}$, 
$\partial_\mu \sigma = - \frac{2 \eta_{\mu\nu} x^\nu}{\eta_{\xi\zeta} x^\xi x^\zeta}$, 
$\partial_\mu \partial_\nu \sigma = - \frac{2 \eta_{\mu\nu}} {\eta_{\xi\zeta} x^\xi x^\zeta} 
+ \frac{4 \eta_{\mu\alpha} x^\alpha \eta_{\nu\beta} x^\beta} {\left(\eta_{\xi\zeta} x^\xi x^\zeta \right)^2}$, 
$\eta^{\alpha\beta} \partial_\alpha \partial_\beta \sigma = 0$, 
when we require $\mathcal{G}_{\mu\nu}$ in (\ref{eq1abcdC}) is proportional to $\eta_{\mu\nu}$, we find 
\begin{align}
\label{coefs}
0=- 4a - b - 8c - 2d\, , \quad 0=- b - d \, .
\end{align}
By combining the above equations with (\ref{abcd}), $0 = a + b + 2c + d$, we find 
\begin{align}
\label{bcd}
b=-8a \, ,\quad c = - \frac{a}{2} \, , \quad d = 8a \, .
\end{align}
and $\mathcal{G}_{\mu\nu}$ in (\ref{eq1abcdC}) has the following form by assuming (\ref{esigma}), 
\begin{align}
\label{eq1C}
\mathcal{G}_{\mu\nu} 
= 12a \eta_{\mu\nu} \eta^{\alpha\beta} \partial_\alpha \sigma \partial_\beta \sigma 
= \frac{48a }{\eta_{\xi\zeta} x^\xi x^\zeta} \eta_{\mu\nu} 
= \frac{24a}{l^2} g_{\mu\nu} \, .
\end{align}
By using (\ref{bcd}), $A_Q$ in (\ref{AQ}) has the following form, 
\begin{align}
\label{AQf}
A_Q=&\, a g^{\alpha\mu} g^{\beta\nu} g^{\gamma\rho} \nabla_\alpha g_{\beta\gamma} \nabla_\mu g_{\nu\rho}
 - 8 a g^{\alpha\mu} g^{\beta\nu} g^{\gamma\rho} \nabla_\alpha g_{\beta\gamma} \nabla_\rho g_{\nu\mu}
 - \frac{a}{2} g^{\alpha\mu} g^{\beta\gamma} g^{\nu\rho} \nabla_\alpha g_{\beta\gamma} \nabla_\mu g_{\nu\rho} \nonumber \\
&\, + 8a g^{\alpha\mu} g^{\beta\gamma} g^{\nu\rho} \nabla_\alpha g_{\beta\gamma} \nabla_\nu g_{\mu\rho} \, .
\end{align}
By using $A_Q$ in (\ref{AQf}), we consider the following action with a cosmological constant $\Lambda$, 
\begin{align}
\label{QJT}
S_{Q-\mathrm{JT}} = \int d^2x \sqrt{-g} \left( A_Q - 2\Lambda \right) \, .
\end{align}
Then by the variation of the action $S_{Q-\mathrm{JT}}$ in (\ref{QJT}), 
we obtain $\mathcal{G}_{\mu\nu} = \Lambda g_{\mu\nu}$. 
Then by choosing $l$ so that $\frac{24a}{l^2} = \Lambda$, 
Eq.~(\ref{eq1C}) shows that we obtain a solution with a constant curvature (\ref{tR}). 
Therefore the action (\ref{QJT}) describes the symmetric teleparallel version of JT gravity. 
That is, the model given by the action (\ref{QJT}) describes the spacetime with constant curvature in the coincident gauge (\ref{Cgauge}). 

The coincident gauge condition (\ref{Cgauge}) is not always compatible with the conformal form of the metric in (\ref{conf}). 
We find the solution of the compatibility as in (\ref{abcd}). 
Furthermore, by considering the conditions that the model has a solution of spacetime with constant scalar curvature as in JT gravity, 
we have shown that there exists such a model as in (\ref{AQf}). 

Finally, let us discuss the ghost problem. 
In the coordinate system (\ref{conf}), $A_Q$ in (\ref{AQf}) has the following form, 
\begin{align}
\label{AQfBB}
A_Q= 8 a \e^{-\sigma} \eta^{\mu\nu} \partial_\mu \sigma \partial_\nu \sigma \, .
\end{align}
Then if we consider the action (\ref{QJT}), a necessary condition to exclude the ghost is that $a$ is negative, $a<0$. 
In order for the conformal mode $\sigma$ to be a canonical scalar field, we require $a=- \frac{1}{16}$. 
Note the condition $a<0$ is necessary but not sufficient. 
In order to check if $a<0$ is a sufficient condition to exclude the ghosts, we need to do a very complicated Hamiltonian analysis. 
This will be done elsewhere.

\section{Summary and Discussions\label{SecVII}}

In summary, we constructed a model describing JT gravity in two dimensions, where the curvature is a constant, 
in the framework of symmetric teleparallel gravity based on a non-metricity tensor in (\ref{non-metricity}). 
Usually, one uses the scalar quantity $Q$ defined by (\ref{non-m scalar}) or (\ref{Q}) in the symmetric teleparallel gravity. 
As given in (\ref{Q}), the non-metricity scalar $Q$ is composed of the bilinear terms of the non-metricity tensor $Q_{\sigma \mu \nu}= \nabla_\sigma g_{\mu \nu}$. 
The combination of the bilinear terms in (\ref{Q}) is not a unique scalar quantity from the viewpoint of covariance because 
the non-metricity tensor $Q_{\sigma \mu \nu}$ is a tensor. 
For an arbitrary combination $A_Q$ in (\ref{AQ}), however, ghosts appear in general. 
Note that $Q$ is a unique combination which includes only gravitational waves as dynamical degrees of freedom and does not include ghosts. 
In two dimensions, however, the propagating gravitational wave does not appear. 
When we consider general scalar quantities as in (\ref{AQ}), any propagating modes might appear although these modes are often ghosts. 
For the general combination in (\ref{AQ}), we investigated the condition that the conformal form of the metric (\ref{conf}) is compatible 
with the coincident gauge (\ref{Cgauge}), where all the connections vanish and found Eq.(\ref{abcd}). 
This condition is rather technical but the non-trivial connections make situations and equations very complicated. 
After that, we consider the conditions where the solution describes the spacetime with constant curvature as in (\ref{esigma}) and (\ref{tR}). It is then found 
(\ref{coefs}) or (\ref{bcd}). 
As a result, we obtain a model (\ref{QJT}) with (\ref{AQf}). 
This model could be regarded as JT gravity in the framework of symmetric teleparallel gravity. 
As clear from (\ref{AQfBB}) when the conformal form of the metric (\ref{conf}) is chosen, the conformal mode $\sigma$ propagates as a scalar mode 
and if $a$ is negative, the scalar mode is not a ghost. 

Then it could be interesting to consider the relation between the obtained formulation of JT gravity and SYK model~\cite{Sachdev:1992fk, Kitaev}. 
In symmetric teleparallel gravity, the concept of geometry is different from that of Einstein gravity and teleparallel gravity using the torsion 
because both the curvature and the torsion vanish. 
Holography based on symmetric teleparallel gravity is not well understood so far. Therefore,
the theory proposed in this paper could be a simple toy model which might clarify the holography or the AdS$_2$/CFT$_1$ correspondence 
in the symmetric teleparallel gravity.

\section*{ACKNOWLEDGEMENTS}

This work was partially supported by the program Unidad de Excelencia Maria de Maeztu CEX2020-001058-M, Spain (S.D.O). 
The work by SDO was also supported by grant No. 24FP-3B021 of the Higher Education and Science Committee of the Ministry of Education, Science, Culture and Sport RA.

\end{document}